\documentclass{elsart}
\usepackage[latin1]{inputenc}
\usepackage{epsfig}
\newcommand{\pde}[1]{\frac{\partial}{\partial {#1}}}

\renewcommand{\Re}{\mathop{\mathrm{Re}}}
\journal{Physica A}
\begin{document}
\begin{frontmatter}
\title{Coupled  three-state oscillators} 
\author{T.~Prager$^{1}$, B. Naundorf$^{2}$, and L. Schimansky-Geier$^{1}$}
\address{$^1$ Institute of
Physics, Humboldt-University of Berlin, Invalidenstr. 110, 10115
Berlin, Germany } 
\address{$^2$ Max-Planck-Institut für
Strömungsforschung and Institut für Nichtlineare Dynamik der
Universität Göttingen, Bunsenstr.~10, 37073 Göttingen, Germany}

\begin{abstract}
  We investigate globally coupled stochastic three-state oscillators,
  which we consider as general models of stochastic excitable
  systems. We compare two situations:in the first case the transitions between
  the three states of each unit $1 \to 2 \to 3 \to 1$ are determined by
  Poissonian waiting time distributions. In the second case only transition $1
  \to 2$ is Poissonian whereas the others are deterministic with a
  fixed delay.  When coupled the second system shows coherent
  oscillations whereas the first remains in a stable stationary state. We show that the
  coherent oscillations are due to a Hopf-bifurcation in the dynamics
  of the occupation probabilities of the discrete states and discuss
  the bifurcation diagram.
\end{abstract}
\end{frontmatter}

\section{Introduction}
Coupled dynamic units are of great importance to understand
cooperative behavior in deterministic dynamical systems. Many
investigations on different levels of description like reaction
diffusion equations, discrete deterministic maps, globally connected
(mean field) or nonlocal models etc. give nowadays a good
understanding of temporal and spatial pattern formation in different
fields of science \cite{Mikhailov}. 

In the case of a coupled stochastic dynamics, stochastic source terms
have to be incorporated in the deterministic dynamics. From
simulations and equations for the moments it was shown that noise may
be the origin of nontrivial behavior leading to noise induced order at
non-vanishing noise intensities \cite{Sancho}. In particular, a new
type of non-equilibrium phase transitions was found and it was shown
that the transmission of signals even benefits from the presence of
noise \cite{zaikin}. 

Principal insight of noise induced behavior has been achieved by
introducing stochastic coupled units with few discrete states.  For
instance, stochastic resonance became a useful application in many
different contexts after the formulation of two-state stochastic
systems with periodically driven rates \cite{McNamara}. Also coupled
chains have explained front motion in fluctuating media and array
enhanced stochastic resonance \cite{Anishchenko}. Noise induced fluxes
in sawtooth like potentials and media have also been considered with
such stochastic systems incorporating a small number of possible
states \cite{Astumian}. 

Most of the present simple discrete systems consider Poissonian
transitions between the $n$ states of a unit. But such approximation,
yielding a Markovian description, require often the introduction of a
large number of variables or states. Whereas random walks are often
considered with different waiting time distributions \cite{Weiss}
investigations of stochastic discrete dynamical units with, for
example, delayed feedback, has started only recently
\cite{Pikovsky,nikitin}. 

In \cite{Naundorf} we presented the occurrence of global oscillations
in coupled three state oscillators with fixed waiting time
\cite{nikitin} as a prototypical model for coupled excitable units.
Here we compare this model with its Poissonian counterpart.  We start
from a common description for stochastic three state dynamics with
general waiting time distributions. Whereas the resulting local
dynamics characterized by its stationary density and power spectrum
behaves similarly in both cases, the coupled cases exhibit a different
behavior.  In the globally coupled network with Poissonian
transitions only stationary states can be encountered due to the high
randomness of the transitions. Contrary, the delayed network shows
coherent oscillations \cite{PikovskyKurthsCR} and synchronization is
found, as observed before in numerical investigations of excitable
systems ~\cite{NedaBarabasi}.
\section{Single three state units}
\begin{center}
\begin{figure}[h]
  \hspace{1cm} \begin{minipage}{4cm}
    \vspace{-0cm}
    \includegraphics[scale=0.9]{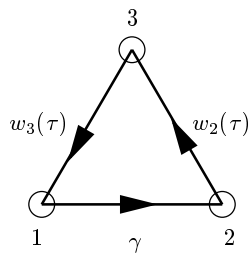}
  \end{minipage}
  \hspace{1cm}
  \begin{minipage}[h]{7cm}
  \caption{\label{Model}A single stochastic three-state unit. The process $1 \to 2$ 
      is Poissonian, while the transitions $2\to 3$ and $3\to 1$
      are are governed by the waiting time distribution $w_2(\tau)$
      and $w_3(\tau)$.}
\end{minipage}
\vspace{-0.5cm}
\end{figure}
\end{center}
Our prototypical model of an excitable system is shown in
Fig.~\ref{Model}.  The unit aims to mimic a single stochastic
excitable system. The transition from state $1 \to 2$ is the
excitation step and occurs with rate $\gamma$. The transitions from $2
\to 3$ and from $3 \to 1$ model the return and take place with
arbitrary waiting time distributions $w_2(\tau)$ and $w_3(\tau)$,
respectively. As output of the unit we assign $f(t)=1$ if the unit is
in state $3$ and $f(t)=0$, otherwise.

First we derive equations which govern the dynamics of the
probabilities $P_i(t)$, $i=1,2,3$, that the system is in state $i$ at
time $t$. To this end we introduce the probability densities
$P_i(t,\tau)$ that the system is in state $i$ at time $t$ \emph{and}
has entered it at time $t-\tau$.  The relation between these
probability densities and the probabilities $P_i(t)$ is given by
$P_i(t)=\int_0^\infty \!d\tau\, P_i(t,\tau)$.  The flow $J_i(t)$ of
probability between state $i$ and $i+1$ can be expressed in terms of
the probability densities $P_i(t,\tau)$ and the waiting time
distributions $w_i(\tau)$.  If this transition is governed by a
Poisson process with rate $\gamma$ the flow is just this rate times
the probability that the particle is in state $i$. Thus, for the
transition $1 \to 2$ $J_1(t)=\gamma P_1(t)$. The probability density
$P_{2}(t,\tau)$ is the flow $J_1$ at time $t-\tau$ times the
probability that the process stays in state $2$ at least the time
$\tau$. The latter is determined by $w_{2}(t)$. In summary,
\begin{equation}
\label{eq1}
P_{2}(t,\tau) = \gamma P_1(t-\tau) \Big[1 - \int_0^\tau d \tau' w_{2}(\tau')\Big]\,.
\end{equation}
By definition we immediately obtain
\begin{equation}
  \label{eq2}
    P_2(t)=\int_0^{\infty}\!d\tau \, \gamma P_1(t-\tau)\Big[1-\int_0^\tau \!d\tau' w_2(\tau') \Big]\\ 
\end{equation}
In case of a general waiting time distribution we introduce a waiting
time dependent conditioned rate $\phi_i(\tau)$ from the relation
\begin{equation}
w_i(\tau)= \phi_i(\tau) \Big[1-\int_0^\tau \!d\tau' w_i(\tau')\Big]\,.
\end{equation}
It gives the rate of escaping from state $i$ after having waited the time $\tau$
and determines the flow between $i$ and $i+1$ 
according to $J_i(t)=\int_0^\infty \!d\tau \phi_i(\tau)
P_i(t,\tau)$. 
Subsuming we obtain
\begin{equation}
P_{i+1}(t) = \int_0^\infty \!d\tau J_i(t-\tau) \Big[1 - \int_0^\tau d \tau' w_{i+1}(\tau')\Big]\,.
\end{equation}
Inserting $P_2(t,\tau)$ from eq.(~\ref{eq1}) into this equation for
$i=2$ finally leads to
\begin{equation}
\label{dyneq}
  P_3(t) = \int_0^{\infty} \!d\tau \int_0^{\infty} \!d\tau' \gamma P_1(t-\tau-\tau')w_2(\tau') \Big[1-\int_0^\tau \!d\tau'' w_3(\tau'')\Big]
\end{equation}
The system of eqs.~(\ref{eq2}) and (\ref{dyneq}) is completed and
becomes closed by the normalization condition
\begin{equation}
\label{norm}
P_1(t)=1-P_2(t)-P_3(t)\,.
\end{equation}
Because of the first Poissonian transition the system is generally
ergodic and possesses one stationary asymptotic solution.

Further information of the oscillator can be extracted from its
spectral properties.  To this end we consider the stochastic process
(output) $f(t)$ as introduced, previously. This process corresponds to
a sequence of pulses whose lengths $\rho$ are distributed according to
$w_\rho(t)=w_3(t)$ and the intervals $\sigma$ between two consequent
pulses are distributed according to $w_\sigma(t)=\int_0^t \!d\tau\, w_1(\tau)
w_2(t-\tau) =\gamma \int_0^t \!d\tau\,\exp(-\gamma \tau) w_2(t-\tau)
$.  The spectrum of $z(t)-\left<z(t)\right>$ can be calculated
employing renewal theory \cite{Stratonovich} and is given by
\begin{eqnarray}\label{spectrum}
  S(\omega)=\frac{4}{\omega^2(\left<\rho\right>+\left<\sigma\right>)}
  \Re\frac{(1-\Theta_\rho(\omega))
    (1-\Theta_\sigma(\omega))}{1-\Theta_\rho(\omega)\Theta_\sigma(\omega)}
\end{eqnarray}
where $\Theta_\rho(\omega):=\left<\exp(i \omega \rho)\right>$ and
$\Theta_\sigma(\omega):=\left<\exp(i \omega \sigma)\right>$ are the
characteristic functions of the random variables $\rho$ and $\sigma$
respectively.

\begin{figure}[!h]
\begin{minipage}{6cm}
  \vspace{0cm}
    \includegraphics[scale=1.0]{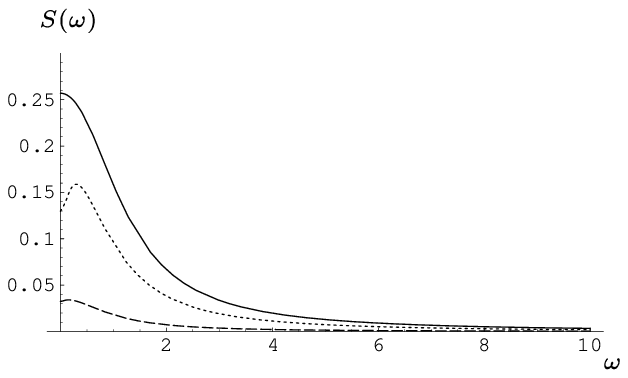}
\end{minipage} 
\hspace{0.8cm}
\begin{minipage}{6cm}
  \includegraphics[scale=1.]{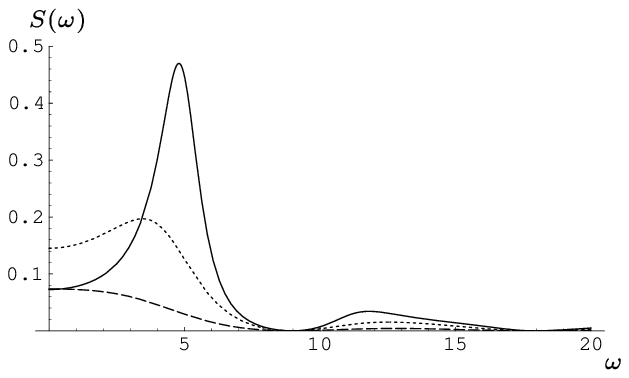}
\end{minipage}
\begin{center}
\caption{The power spectrum for a single stochastic oscillator which
  Poissonian (left) and deterministic waiting time distributions
  (right).  Left: $\gamma=0.1,\gamma_3=1$.  Solid: $\gamma_2=1$,
  dotted: $\gamma_2=0.1$. dashed:$\gamma_2=0.01$. Right:$\tau_2=0.3$
  and $\tau_3=0.7$. Solid: $\gamma=2.0$, dotted:$\gamma= 0.5$, dashed:
  $\gamma=0.1$.
  \label{figure4a}}
\end{center}
\end{figure}

Let us now consider the case where all three transitions are governed
by Poisson processes, i.e. $w_2(\tau)=\gamma_2 \exp(-\gamma_2 \tau)$
and $w_3(\tau)=\gamma_3 \exp(-\gamma_3 \tau)$.  After taking the
derivative of eq.  (\ref{eq2}) with respect to $t$ we obtain
\begin{eqnarray*}
     \dot{P}_2(t)&=&-\int_0^{\infty} \!d\tau\, \Big[\pde{\tau}\gamma P_1(t-\tau)\Big]
     \exp(-\gamma_2 \tau)\,=\,-\gamma_2 P_2(t)+\gamma P_1(t)
\end{eqnarray*}
A similar calculation can be done for $P_3(t)$, leading to
\begin{eqnarray*}
  \dot{P}_3(t)&=&-\gamma_3 P_3(t)+\gamma_2 P_2(t),
\end{eqnarray*}
i.e. one retrieves the expected rate dynamics.  To determine the
corresponding spectrum as defined in the previous section we use
$w_\rho(t)=\gamma_3 \exp(-\gamma_3 \rho)$ and
$w_\sigma(t)=\frac{\gamma\gamma_2}{\gamma-\gamma_2}(\exp(-\gamma_2
t)-\exp(\gamma t))$ which gives (see Fig.~(\ref{figure4a}))
\begin{eqnarray*}
  S(\omega)=\frac{4((\gamma+\gamma_2)^2+\omega^2)}
  {(\frac{1}{\gamma}+\frac{1}{\gamma_2}+\frac{1}{\gamma_3})((\gamma\gamma_2+\gamma_2\gamma_3+\gamma_3\gamma)^2+\omega^2(\gamma^2+\gamma_2^2+\gamma_3^2)+\omega^4)}.
\end{eqnarray*}

Instead of a fixed rate for the transition $2\to 3$ and $3\to 1$ we
now deal with the case of a fixed waiting time. The corresponding
waiting time distributions are given by
$w_2(\tau)=\delta(\tau-\tau_2)$ and $w_3(\tau)=\delta(\tau-\tau_3)$.
Inserting these into eqs.~(\ref{dyneq}) and (\ref{eq2}) leads to
\begin{eqnarray*}
  P_2(t)&=&\int_0^{\tau_2} \!d\tau\, \gamma P_1(t-\tau),\quad
  P_3(t)=\int_0^{\tau_3}\!d\tau\, 
  \gamma P_1(t-\tau-\tau_2) 
\end{eqnarray*}
supplemented by normalization, eq.~(\ref{norm}). The calculation of
the power spectrum is done analogously to the Markovian case, finally
giving
\begin{eqnarray*}
\hspace{-0.5cm}  S(\omega)
  &=& \frac{2\cos(\omega
    \tau_3)-2}{\big(\frac{1}{\gamma}+\tau_2+\tau_3\big)\big(\omega^2+\gamma^2\big)
    \big(1+\frac{\gamma}{\sqrt{\omega^2+\gamma^2}}
    \sin((\tau_2+\tau_3)\omega+\phi)\big)},\quad \tan \phi=\frac{\omega}{\gamma}.
\end{eqnarray*}
With $\tau_2$ fixed to 0.3 and $\tau_3$ fixed to 0.7
the spectra for different values of the rate $\gamma$  
are presented in figure \ref{figure4a}.

The spectra do not differ too much from the Poissonian case. In both
cases it is possible to have a maximum in the spectrum at 
frequencies different from zero.

\section{Globally coupled three state units}
We now consider a set of $N$ three state oscillators as described in
the previous section.  A global coupling between the single units is
introduced by letting the rate $\gamma$, which governs the transition
$1 \to 2$, depend on the common output
\begin{eqnarray}\label{deponstate}
  f(t)=\frac{1}{N} \sum_{i=1}^N f_i(t)
\end{eqnarray}
where $f_i(t)$ is now the output of a single unit. 

In the limit $N\to\infty$ the state of the total system can be
described by the ensemble averaged occupation probabilities $P_i(t)$
that a single oscillator is in state $i$ at time $t$. Since $f(t)$
converges to the occupation probability of the third state, $\gamma$
becomes a function of $P_3(t)$.

Equations which govern the dynamics of the ensemble averaged
occupation probabilities are derived in the same manner as
eqs.~ (\ref{dyneq}) and  (\ref{eq2}) except that one has to take care that the rate
$\gamma$ is no longer a constant but is dependent on $P_3(t)$. One
eventually arrives at
\begin{eqnarray}\label{dyneq2}
  \hspace{-0.5cm}P_2(t)&=&\int_0^{\infty} \!d\tau\,\gamma\big(P_3(t-\tau)\big)
  P_1(t-\tau)\Big[1-\int_0^\tau \!d\tau'\,
  w_2(\tau')\Big]\\ 
  \hspace{-0.5cm}P_3(t)&=&\int_0^{\infty}\!d\tau \int_0^{\infty}\!d\tau'\, 
  \gamma\big(P_3(t-\tau-\tau')\big) P_1(t-\tau-\tau')w_2(\tau')
\Big[1-\int_0^\tau \!d\tau''\, w_3(\tau'')\Big]\nonumber
\end{eqnarray}
and normalization, eq.(\ref{norm}). These equations are now nonlinear
in the occupation probabilities due to the dependency of $\gamma$ on
the state of the system.

The system dynamics of the coupled system can be understood by
analyzing the stationary solutions and its stability properties. The
stationary solutions $P_1^*, P_2^*$ and $P_3^*$ are defined by setting
$P_i(t)=\mbox{const.}$ in eq.~(\ref{dyneq2}). Using
$\int_0^{\infty}(1-\int_0^\tau \!d\tau'\,w_{2/3}(\tau')
)\!d\tau=\int_0^\infty \!d\tau\, \tau w_{2/3}(\tau) =:\bar w_{2/3}$,
which is the mean waiting time in state $2$ or $3$, respectively, we
obtain implicitly
\begin{equation}\label{statsol}
P_1^*=\frac{\gamma^{-1}(P_3^*)}{T},\;~~~P_2^*=\frac{\bar w_2 }{T},\;~~~P_3^*=\frac{\bar w_3 }{T}.
\end{equation}
with $T=\gamma^{-1}(P_3^*)+\bar w_2+\bar w_3$ being the mean time of
one cycle. In contrast to the uncoupled case, where  exactly one
asymptotic stationary solution exists, the asymptotic behavior may now
be more complex due to the nonlinearity of eqs.~(\ref{dyneq2}).
Depending on the chosen coupling function there can be several
not necessarily stable stationary solutions.

The local stability of a stationary solution can be derived by
inserting the Ansatz $P_i(t)=P_i^*+A_i \exp(\lambda t),\; \lambda \ne
0,$ into the linearized dynamical equations. Therefrom we obtain the
characteristic equation
\begin{eqnarray}\label{chareq}
 &&0=1+ \gamma(P_3^*)
 \int_0^{\infty}\!d\tau\, e^{-\lambda\tau}(1-\int_0^\tau \!d\tau'w_2(\tau'))\\
  &&+\,\big(\gamma(P_3^*)-
  \gamma'(P_3^*)P_1^*\big)\int_0^{\infty}\!d\tau \Big[
  e^{-\lambda\tau}(1-\int_0^\tau \!d\tau'  w_3(\tau')) \Big]
  \int_0^{\infty}\!\!d\tau\, e^{-\lambda\tau}w_2(\tau)\nonumber
\end{eqnarray}

In the case with Poissonian transitions the mean waiting times in
state $2$ and $3$ are given by $\bar w_2=1/\gamma_2$ and $\bar
w_3=1/\gamma_3$.  There may be several stable and unstable stationary
solutions depending on the chosen coupling function.  
Eq.~(\ref{chareq}) 
corresponds to the characteristic
equation of a two dimensional dynamical system:
\begin{eqnarray*}
  \lambda^2+\lambda(\gamma(P_3^*)+\gamma_2+\gamma_3)+\gamma(P_3^*)\gamma_2+
  \gamma_2\gamma_3 + \gamma_3 \gamma(P_3^*)- \gamma_2 \gamma'(P_3^*)
  P_1^*=0.
\end{eqnarray*}
In order to decide whether there exist limit cycles 
we calculate the divergence of the flow in the phase space which is
$-(\gamma(P_3)+\gamma_2+\gamma_3)$. As the rates $\gamma(P_3)$,
$\gamma_2$ and $\gamma_3$ are always
positive the divergence is everywhere negative. Therefore no limit
cycles exist, i.e. the system shows no coherent oscillations.

For fixed waiting times $\tau_2$ and $\tau_3$
the dynamics of the occupation probabilities is reduced to
\begin{eqnarray} \label{ddeqs}
  P_{2}(t)  &=&  \int_{0}^{\tau_2}\!d\tau\,\gamma (P_{3}(t-\tau))P_{1}(t-\tau)\,
  ,\\
  P_{3}(t)  &=&  \int _{0}^{\tau_3}\!d\tau\,
  \gamma (P_{3}(t-\tau-\tau_2))P_{1}(t-\tau-\tau_2)\,,\quad
   P_{1}(t)  =  1-P_{2}(t)-P_{3}(t)\nonumber
\end{eqnarray}
The stationary solutions are given by eq.~(\ref{statsol})
with $\bar w_2=\tau_2$ and $\bar w_3=\tau_3$. The characteristic
equation is easily derived from the general
eq.~(\ref{chareq}) to be
\begin{eqnarray}
  \gamma(P_3^*)(e^{-\lambda (\tau_2 +\tau_3)}-1)-
  \gamma'(P_3^*)
  P_1^*(e^{-\lambda (\tau_2+\tau_3)}-e^{-\lambda \tau_2})-\lambda=0.
\end{eqnarray}
Such characteristic equations are typical for dynamical systems
involving fixed delays \cite{delay}.
\section{Sigmoidal coupling}
In the following we review the behavior of the coupled system
for a sigmoidal rate function as presented in \cite{Naundorf}
\begin{eqnarray*}
  \gamma(P_3(t))=\gamma_0\,\left(1+\Delta\tanh\left[-\frac{f(t)-f^{*}}{
2\sigma} \right]\right).
\end{eqnarray*}
For this rate function, there exists exactly one stationary solution
of the system dynamics. The coupling strength is mediated by $\sigma$:
The smaller $\sigma$ the stronger is the dependence of the rate on
$P_3(t)$. For small $\sigma$ the rate changes rapidly from
$\gamma_1:=\gamma_0(1+\Delta)$ to $\gamma_2:=\gamma_0(1-\Delta)$ if
the value of $P_3(t)$ crosses $f_0$. For large values of $\sigma$ the
rate is nearly independent of $P_3(t)$ and equal to $\gamma_0$.

In case of a Poissonian waiting time distribution this stationary
solution is necessarily stable as shown in the last section. 
The high randomness in the transitions does
not allow the excitation of oscillating behavior, i.e 
the coupled system does not differ substantially from the uncoupled case.

For fixed waiting times, however, the behavior changes due to the
coupling.
\begin{figure}[!h]
  \begin{center}
    \includegraphics[scale=0.5]{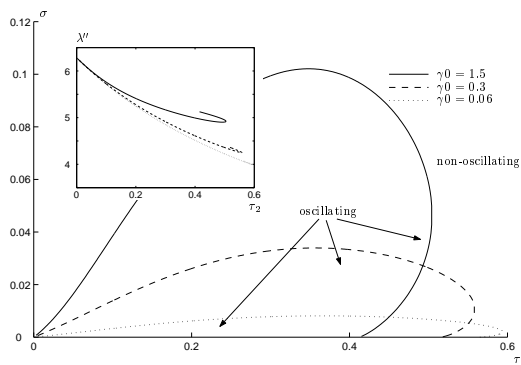}
\caption{Solutions of the characteristic equation with $\Re \lambda=0$ for
  different values of $\gamma_0$.  The parameter $\Delta$ is fixed to
  2/3, $f_0$ is chosen to be the stationary solution $P_3^*$ at
  $\tau_2=0.3$, which is about 0.42, 0.16 and 0.04 respectively.  The
  inset shows the corresponding frequency at the bifurcation.
\label{figure4}}
\end{center}
\end{figure}
For each value of $\gamma_0$ there is a finite region in which the
ensemble oscillates. This region grows with increasing values of
$\gamma_0$. If $\sigma$ is chosen too large, coherent behavior cannot
be observed.  Interestingly, there exist values of $\tau_2$ where
$\sigma$ has to exceed some finite value to trigger the oscillations.
We have verified this transition between non-oscillating, oscillating and again
non-oscillating behavior in numerical
simulations.

\begin{figure}[!h] 
\begin{minipage}{6cm}
  \includegraphics[scale=0.5]{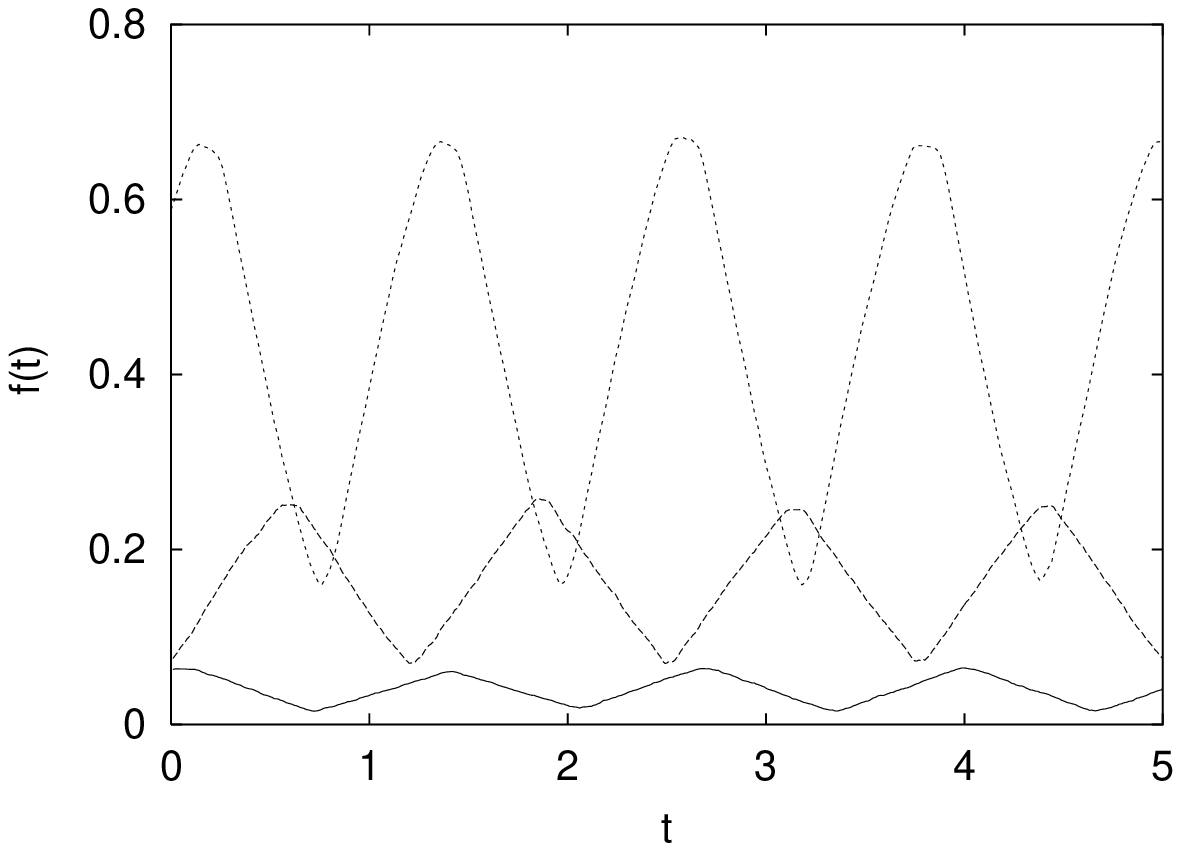}
\end{minipage} 
\hspace{0.3cm}
\begin{minipage}{6cm}
  \includegraphics[scale=0.5]{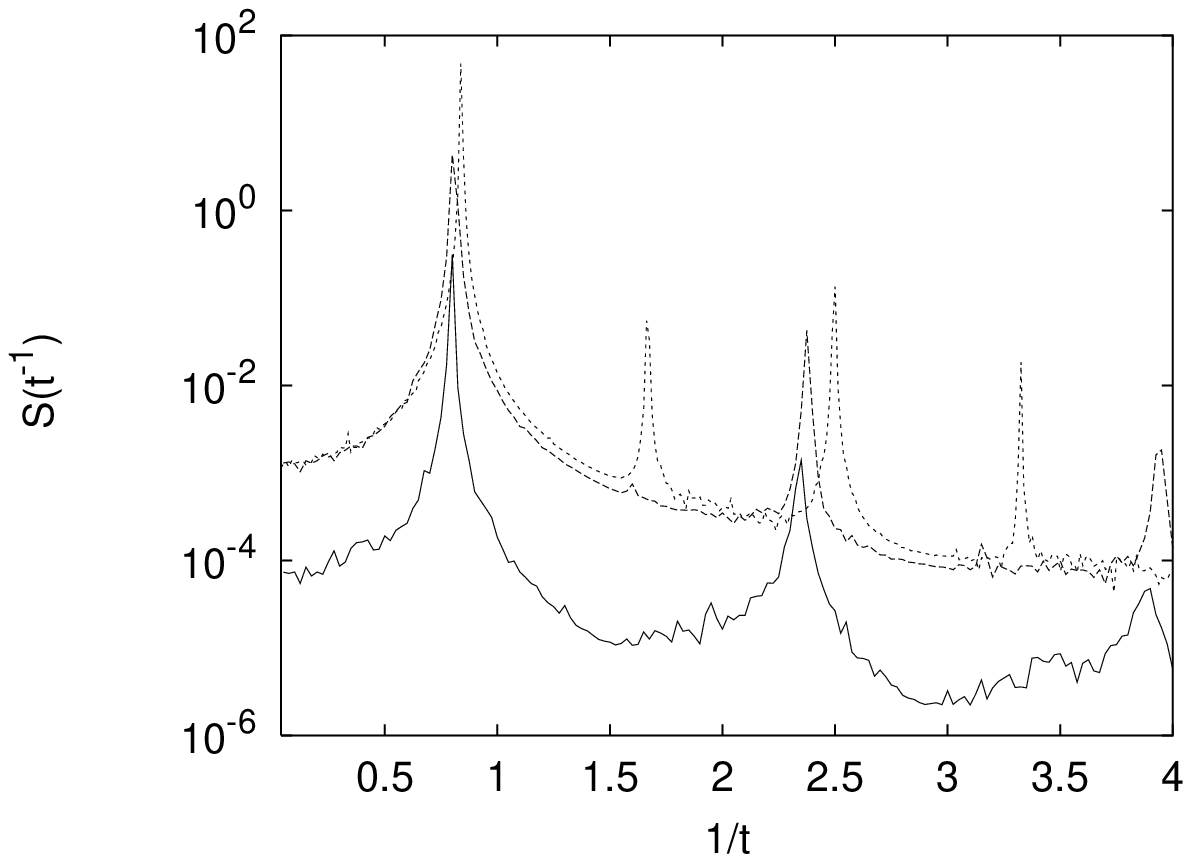}
\end{minipage}
\begin{center}
\caption{Left: Behavior of the ensemble for different values 
$\gamma_0$ and $f_0$ with  $\sigma$ fixed to 0.001 and $\Delta$ to 2/3
and $\tau_2$ to 0.3.
\noindent Solid: $\gamma_0=0.06, f_0=0.04$
\noindent Dashed: $\gamma_0=0.3, f_0=0.16$
\noindent Dotted: $\gamma_0=1.5, f_0=0.42$ 
The simulations have been performed for a system consisting of 10000
units.  Right: The corresponding spectra of the common output $f(t)$ 
\label{figure5}}
\end{center}
\end{figure}

One may ask, whether there is some connection between the oscillatory
behavior of a single oscillator and the oscillatory behavior of the
total system. To this end we have carried out simulations with the
sigmoidal rate function, where $\gamma_0$ has been chosen to be
$0.06$, $0.3$ and $1.5$ respectively.  In the first case the rate lies
always in the non oscillating region of a single oscillator, in the
second case, it switches between the oscillating and non oscillating
region, and in the third case it is always in the oscillating region.
$\sigma$ has been fixed to $0.001$ and the values of $f_0$ are chosen
to be approximately equal to the stationary solution $P_3^*$. The
results are shown in figure~\ref{figure5}.
In each case we observe coherent oscillations of the total system,
although the amplitude of the oscillations depends strongly on the
chosen rate.  Interestingly however, the frequency of the oscillations
is nearly independent on $\gamma_0$ although the mean time of
one cycle of a single oscillator strongly varies with $\gamma_0$.
Taking into account that the mean waiting time of a Poisson process
with rate $\gamma$ is $1/\gamma$ the mean time for one cycle of a
single oscillator lies between $1+1/(\gamma_0(1+\Delta))$ and
$1+1/(\gamma_0(1-\Delta))$. This corresponds to frequencies between
approximately 0.02 and 0.09 for $\gamma_0=0.06$, 0.09 and 0.3 for
$\gamma_0=0.3$ and 0.3 and 0.7 for $\gamma_0=1.5$.  Obviously the
frequencies of the coherent oscillations one observes do not
correspond to these frequencies, especially in the first case they are
even of a different order of magnitude.  However there is a good
agreement between the observed frequencies and the frequencies at the
bifurcation, which are shown in the inset of Fig.~\ref{figure4}.

\section{Conclusion}
We have investigated stochastic three state oscillators, which we
propose as a model for excitable systems driven by noise. First we
considered the behavior of the single units and inspected their
spectral properties for the case of Poissonian and deterministic
transitions. It turns out that the process with deterministic
transitions is a good simple model describing excitable dynamics with
rich spectral behavior.  When coupled globally both models behave
quite differently.  While in the Poissonian case there are no
oscillating solutions, in the case with fixed waiting times the
coupling leads to coherent oscillations of the network. A stability
analysis reveals, that these oscillations are due to a Hopf
bifurcation in the dynamics of the occupation probabilities.
Interestingly, the global oscillations occur even for units which have
maximal power at zero frequency. It is the common situation of noise
induced phase transitions, that coupling might induce ordered behavior
despite the fact that the constituting elements involve a lot of
randomness \cite{Sancho,zaikin,Anishchenko}.

This work was supported by DFG-Sfb 555. We thank B. Lindner and M.
Zaks for help and fruitful comments.


\begin{thebibliography}{10}   
\bibitem{Mikhailov} A. S. Mikhailov, {\em Foundations of Synergetics
    I}(Springer, Berlin-Heidelberg-New York, 1990)
\bibitem{Sancho} C. van den Broeck , J. M. R. Parrondo, and R. Toral,
  Phys. Rev. Lett. 73, 3395 (1994); J. García-Ojalvo and J. M.
  Sancho,{\em Noise in Spatially Extended Systems} (Springer, New
  York, 1999).
\bibitem{zaikin} A. A. Zaikin, J. García-Ojalvo, L. Schimansky-Geier,
  and J. Kurths Phys. Rev. Lett. 88, 010601 (2002).
\bibitem{McNamara} B. McNamara and K. Wiesenfeld, Phys. Rev A
  \textbf{39}, 4854 (1989).
\bibitem{Anishchenko} V. Anishchenko, et al. {\em Chaotic and
    Stochastic Processes in Dynamic Systems}( Springer,
  Berlin-Heidelberg-New York, 2002).
\bibitem{Astumian}R. D. Astumian, Science \textbf{276}, 917 (1997);
  L. Schimansky-Geier, M. Kschischo, T. Fricke Phys. Rev. Lett.  79,
  3335 (1997); J. A. Freund, and L. Schimansky-Geier Phys. Rev. E 60,
  1304 -1309 (1999).
\bibitem{Weiss} G.  H.  Weiss, Aspects and Applications of the Random
  Walk (North-Holland, Amsterdam, 1994)
\bibitem{Pikovsky} L.  S.  Tsimring and A.  Pikovsky Phys.  Rev.
  Lett. 87, 250602 (2001); L.~Q.~Zhou, X.~Jia, and Q.~Ouyang,
  Phys.~Rev.~Lett.~\textbf{88}, 138301 (2002)
\bibitem{nikitin} A.~Nikitin, Z.~N{\'e}da, and T.~Vicsek
  Phys.~Rev.~Lett.  \textbf{87}, 024101 (2001).
\bibitem{Naundorf} B. Naundorf, T. Prager and L. Schimansky-Geier, submitted for publication
\bibitem{PikovskyKurthsCR} A.~Pikovsky, J.~Kurths,
  Phys.~Rev.~Lett.~\textbf{78}, 775 (1997); B.  Lindner and L.
  Schimansky-Geier, Phys. Rev E \textbf{61}, 6103-6110 (2000).
\bibitem{NedaBarabasi}C. Kurrer and K. Schulten, Phys. Rev. E
  \textbf{51}, 6213 (1995); H.  Hempel, L. Schimansky-Geier, and J.
  Garcia-Ojalvo Phys. Rev. Lett.  \textbf{82}, 3713-3716 (1999); A.
  Neiman, L.  Schimansky-Geier, A. Cornell-Bell, and F. Moss Phys.
  Rev. Lett. \textbf{83}, 4896-4899 (1999).
\bibitem{Stratonovich} R. L. Stratonovich, Topics in the Theory of
  Random Noise I (Gordon and Breach, 1962) p. 175.; B.  Lindner and L.
  Schimansky-Geier, Phys. Rev E \textbf{60} , 7270-7276 (1999).
\bibitem{delay} G. T. Gurija and M. A. Lifshits Z. Phys B \textbf{47}, 71 (1982); L. Schimansky-Geier, Ch. Z\"ulicke and E. Sch\"oll, Physica A \textbf{188}, 436 (1992).

\end{thebibliography}
\end{document}